%% file: paper.tex
\documentclass[letterpaper]{article} 
\usepackage{aaai2027}  
\usepackage[hyphens]{url}  
\usepackage{graphicx} 
\usepackage{natbib}  
\usepackage{caption} 
\usepackage{algorithm}
\usepackage{algorithmic}

\usepackage{newfloat}
\usepackage{listings}
\DeclareCaptionStyle{ruled}{labelfont=normalfont,labelsep=colon,strut=off} 
\floatstyle{ruled}
\newfloat{listing}{tb}{lst}{}
\floatname{listing}{Listing}

\usepackage{tikz}
\usepackage{amsmath}
\usepackage{filecontents}

\usepackage{multirow}
\usepackage{tabularx, booktabs}
\usepackage{xcolor, colortbl}
\usepackage{soul}
\usepackage{xspace}
\usepackage{makecell}
\usepackage{subfig}
\usepackage{xurl}
\usepackage{footmisc}
\newcommand{\K}{\textsc{k}\xspace}
\newcommand{\M}{\textsc{m}\xspace}
\newcommand{\B}{\textsc{b}\xspace}
\newcommand{\cse}{}
\newcommand{\health}{}

\newcolumntype{Y}{>{\centering\arraybackslash}X}
\newcolumntype{R}{>{\raggedleft\arraybackslash}X}
 \newcommand{\Hquad}{\hspace{0.5em}}

\title{Something to Talk About: Social Media as a Lens\\on Healthcare Ransomware Events}
\author{Seoyoung Kweon\cse, Paul Chung\cse, Isabel Straw\cse, Christian J.\ Dameff\health, Jeffrey L.\ Tully\health,\\ Stefan Savage\cse, Geoffrey M.\ Voelker\cse, Deepak Kumar\cse \\[0.1in]
  }
\affiliations{

  University of California, San Diego\\
    \{pekweon, yic144, isstraw, ssavage, voelker, kumarde\}@ucsd.edu\\
    \{cdameff, jtully\}@health.ucsd.edu
}

\begin{document}

\maketitle

\begin{abstract}
In the modern era, ransomware attacks on critical healthcare
organizations---like hospitals and insurers---are frequent, with impacts
ranging from leaked private health information all the way to serious
disruptions of urgent, time-sensitive clinical operations. Unfortunately,
legal and economic incentives make it uncommon for hospitals to share basic
information about these attacks, their scope, and the downstream impacts to
patient care. In this paper, we explore the use of public social media
posts---authored both by hospitals and by individuals---to garner more
detailed insights about these attacks and their effects. We collect
1,628 Facebook and Reddit posts from 2018--2024, design and evaluate techniques
to match post data to 212 ground-truth ransomware attacks, and conduct
quantitative and qualitative analyses that explore the impacts such attacks have
on critical hospital infrastructure, patient care, and providers. We conclude
by discussing the promise and limitations of leveraging social data to
study the impact of ransomware attacks and highlight areas of future research. 
\end{abstract}

\input{introduction} 
\input{related-works} 
\input{methodology} 
\input{analysis} 
\input{case-study} 
\input{discussion} 
\input{acks}

\bibliography{refs}


\input{checklist}

\appendix

\input{real_appendix}

\end{document}

%% file: introduction.tex
\section{Introduction}
Since 2016, U.S.\ hospitals have become an increasingly popular target for
ransomware attacks, rising from  58~reported attacks in 2018 to
284~in 2024~\cite{9509877,comparitech}. The average recovery cost per hospital,
excluding ransom payments, reached \$2.57 million in 2024~\cite{hipaa}.
Beyond financial harm, ransomware attacks can cause internal system disruption, either
by encrypting critical patient medical data or by forcing hospitals to take
systems offline to contain damage after initial exploitation~\cite{Cheon}.
These disruptions hinder a hospital's ability to operate on schedule,
critically affecting patient care and staff capabilities alike. Indeed,
multiple empirical studies have established direct impact on patient outcomes
at not just hospitals targeted by ransomware but
also nearby hospitals who must suddenly accommodate emergency
diversions~\cite{mcglave23,neprash24,dameff23,pham24}. As the American Hospital Association put it,
ransomware attacks ``are threat-to-life crimes because they directly threaten
a hospital's ability to provide patient care''~\cite{aharansomware}.


Unfortunately, understanding the scope and impact of ransomware attacks on
hospital systems remains an outstanding challenge. Most U.S.\ hospitals are loath
to disclose details due to legal liability and brand impact, and, even
in the case of subsequent civil action, such documentation is commonly
sealed~\cite{schwarcz2022privilege,mund2021privilege}.
While the U.S. Department of Health and Human Services maintains a public
portal on some known attacks, details about each attack are inconsistent,
descriptions are limited in scope to breaches of private health information,
and ultimately such reports do not provide the depth needed to understand
downstream impact to patients and hospitals~\cite{munoz2024thematic}.

In this paper, we seek to bridge this gap by exploring an alternative data
source for understanding hospital ransomware impact: social media. In
particular, we explore two distinct channels: a) how hospitals themselves
communicate about such incidents via their official presence on Facebook, and
b) how individuals, both patients and employees, communicate about ransomware
events on Reddit. We collate a ``ground truth'' dataset (from collections
of public reports of hospital ransomware attacks), build and evaluate a
methodology to match ransomware-related content to each attack, and leverage
309~Facebook posts and 1,319~Reddit posts to study the impact of hospital
ransomware attacks over a six-year period from 2018--2024.

We observe significant discussion about ransomware attacks online, with posts
matching 212~ground truth attacks. While Facebook covers fewer attacks than
Reddit (56 vs.\ 180), Facebook posts offer deeper insights into smaller,
individual hospitals and clinics, compared to Reddit posts which are more
focused on larger attacks on hospital systems and third-parties (e.g.,
insurance companies). Facebook posts typically appear faster (median
2.4~days to notification) than Reddit (median 29~days to notification), and seek
to primarily alert patients about impacted patient services (e.g., downtime to
electronic health records, etc.).  In contrast, Reddit discussion is much more
prolonged (averaging 45~days to Facebook's 7.7~days) and provides critical
patient and provider details about recovery post-attack. 

Next, we conduct a qualitative analysis of all post data, highlighting seven
distinct themes of downstream impacts to computer/network infrastructure,
pharmacy services, hospital-facing infrastructure, patient-facing
infrastructure, critical care requiring patient diversion, private health data
via data breaches, and patient, provider, and staff mental health. We
show that Facebook and Reddit offer different lenses to examining impact:
for example, Facebook data more frequently discusses IT infrastructure
impact (78\% of posts vs.\ 30\% of Reddit posts), whereas Reddit data highlights
the financial impact of ransomware on individual patients and providers (31\% of posts
vs.\ 5.6\% of posts).

Altogether, our quantitative and qualitative results present insight into
the myriad deleterious impacts of ransomware in healthcare, and furthermore
demonstrate how Facebook and Reddit can serve as complementary measurement
vantage points. We conclude with a discussion of our results, and offer suggestions
on how insights from social media might be used to both remedy and improve
hospital resilience in the face of future attacks.
We release our detection code, labeled post
datasets, and our full qualitative codebook.\footnote{\url{https://github.com/sykweon/artifact_ransomware_social_media}}




%% file: related-works.tex
\section{Related Work}

This study relates to prior work on ransomware in general, ransomware
attacks on hospitals specifically, and the use of social media data to
study the impact of real-world events.

\subsection{Ransomware Tracking}
Considerable research has tracked the landscape of ransomware attacks and
archived historical incidents. Most recently, \citeauthor{sarabiransomware}
(\citeyear{sarabiransomware}) analyzed the longitudinal patterns of ransomware
using a public news media dataset, focusing on victim behaviors in ransom
payments. Significant prior attention has also tracked ransomware through the
lens of payments. For example, \citeauthor{simoiu2019told}
(\citeyear{simoiu2019told}) analyzed payment methods and victim behaviors with a
focus on individual security practices, and \citeauthor{huang2018tracking}
(\citeyear{huang2018tracking}) developed an end-to-end method in payment
activities to track ransomware over a two-year period of time. Beyond academic
work, several large-scale projects also identify and archive ransomware
incidents through news reports and government datasets. Examples include
Hackmageddon, Privacy Rights Clearing House, Critical Infrastructure Ransomware
Attacks, and the Map of Worldwide Ransomware Attacks, two of which we use in this
paper~\cite{hackmegaddon,prc-data,cira,comparitech}.

Beyond longitudinal analyses, \citeauthor{zhang2018aftermath}
(\citeyear{zhang2018aftermath}) examined a single, large-scale ransomware attack
on an academic institution. Their work assessed ransomware impact beyond payment
and focused on the experiences of associated individuals, which the researchers
describe as ``secondary costs''. From the secondary costs observed, they
provided an attack response plan for large organizations.  Our work specifically
explores the healthcare ecosystem and how social media data can help to unearth
such ``secondary costs'' in health and hospital-specific contexts.

\subsection{Ransomware Attacks in Hospitals}

Several efforts have also examined ransomware specifically in a healthcare context. \citeauthor{minnaar2021cyberattacks} (\citeyear{minnaar2021cyberattacks}), \citeauthor{neprash2022trends} (\citeyear{neprash2022trends}), and \citeauthor{spence2018ransomware} (\citeyear{spence2018ransomware}) all analyzed the long-term rise in frequency of ransomware targeting healthcare, with Minnaar et~al. and Neprash et~al. specifically focusing on changes in attack patterns across different organizational types. Notably, much of this work relies on self-reported survey data from hospitals, rather than leveraging other public sources (e.g., social media). \citeauthor{spence2018ransomware} (\citeyear{spence2018ransomware}) further analyzed the financial and reputational impacts of ransomware on victim organizations through interviews with a healthcare law expert and a hospital chief information officer. Based on these insights, they proposed prevention and response plans to mitigate the effects of such attacks. Other work focuses on the organizational impact of ransomware through the lens of compliance (e.g., HIPAA) and workplace culture~\cite{farringer2016send}. Most of these efforts focused on the impact to hospitals or hospital systems in isolation. However, \citeauthor{dameff23} (\citeyear{dameff23}) explored the ``spillover effect'' of ransomware on \textit{nearby} hospitals when patients are diverted, and the immediate impact to patients as a result. Similar to our qualitative analysis, \citeauthor{munoz2024thematic} (\citeyear{munoz2024thematic}) conducted a thematic analysis of 76~instances of hospital-focused ransomware collected by the U.S.\ Department of Health and Human Services to examine the impact on both organizations and patients.


Although these efforts align with our study of identifying and
analyzing ransomware impacts using public datasets, our study differs
in two key ways. First, we leverage social media data, capturing the
perspectives of both organizations and individuals on ransomware
impact.  No research has leveraged social media communication from
organizations, patients, and providers as a primary data source in
understanding the impact of ransomware in practice.  Second, rather
than focus on financial or reputational outcomes, we use rich
qualitative evidence to highlight the direct and indirect impacts and
consequences of ransomware events in healthcare settings.


\subsection{Leveraging Public Data}
Leveraging social media data is a well-established approach for studying the impact of real-world crisis events. Beyond ransomware analysis, \citeauthor{andalibi2016understanding} (\citeyear{andalibi2016understanding}) and \citeauthor{wei2024understanding} (\citeyear{wei2024understanding}) examined the aftermath of e-crime and abuse using social media data, applying qualitative analysis to Reddit discussions to better understand help-seeking behaviors.
Recently, \citeauthor{saha2025mental} (\citeyear{saha2025mental}) used anxiety and depression related posts in specific subcommunities of Reddit to understand the impact of the COVID-19 pandemic on the mental health of college students.

Other work leverages public datasets such as news or social media for event detection. \citeauthor{satyapanich2020casie} (\citeyear{satyapanich2020casie}) and \citeauthor{mittal2016cybertwitter} (\citeyear{mittal2016cybertwitter}) developed detection systems for cybersecurity-related events, with Satyapanich~et~al. specifically extracting details such as attackers, victims, and tools from news articles. Mittal et al. demonstrated how social media can be used as an early warning system for cybersecurity issues.
Additional studies have built detection systems for other domains, including natural disasters and emergencies \cite{hasan2016twitternews,paul2020outage,otal2024llm}.

%% file: methodology.tex
\section{Methodology}
We collected two different types of social media data in our work: 
Facebook discussions (posts and comments)
from official hospital accounts and Reddit discussions
related to ransomware attacks against hospitals. Facebook \textit{posts} represent
official communication from targets of ransomware, while Facebook comments and Reddit discussions
represent reactions from other affected parties (e.g., providers, patients). In
this section, we detail each step we took to identify and analyze relevant
social conversation about ransomware attacks against hospitals.
Table~\ref{tab:data_summary} summarizes the data used in this study.

\begin{table}[t]
\begin{center}
  \begin{tabular}{@{}llrr@{}}
  \toprule
  \textbf{Data Source} & \multicolumn{2}{p{1.5in}}{\textbf{Use Case} \hfill \textbf{\# Entries}} & \textbf{\# Attacks} \\
  \midrule
    Temple & Ground Truth & 238 & 238 \\
    Comparitech & Ground Truth & 851 & 851 \\
    ID Theft Center & Ground Truth & 4,934 & 520 \\
    CA DoJ & Ground Truth & 3,943 & 96 \\
    \midrule
    Total Unique & & & 938 \\
    & & & \\
    \midrule
    Facebook & Impact Analysis & 309 & 56 \\
    Reddit & Impact Analysis & 1,319 & 180 \\
    \midrule
    Total Unique & & & 212 \\
    & & & \\
    \midrule
    AHA Dataset & Hospital Metadata & 6,225 & \textsc{n/a} \\
     \bottomrule
\end{tabular}
\caption{\label{tab:data_summary} Summary of datasets, including
  source, use case, number of entries, and number of associated
  attacks. Some show fewer attacks than entries because some breaches
  were not ransomware-related.  We extracted data from all sources
  covering January 2018 to December 2024, except for the AHA
  dataset which was published in 2024.}
\end{center}
\end{table}

\subsection{Collecting Known Ransomware Attacks}
\label{sec:collecting_known_ransomware_attacks}

We first compiled a ground truth dataset of ransomware attacks using four
databases: the Breach Alert dataset from \cite{itrc}, the Critical
Infrastructure Ransomware Attacks (CIRA) dataset from Temple
University~\cite{cira}, the Ransomware Attack Map from
Comparitech~\cite{comparitech}, and the Data Security Breach list from
\cite{gov}. We filtered each for ransomware incidents affecting U.S.\ healthcare
organizations covering January~2018 and December~2024.
Because each dataset uses a bespoke naming convention, we consolidated
ransomware events using the names provided by the survey from the American
Hospital Association (AHA), a database that comprises the metadata of
6,225~hospitals in the U.S\@. Our final corpus of ransomware contains 938~events
from January 2018 to December 2024 affecting 897 hospitals, hospital-systems,
and healthcare companies.






\subsection{Identifying Hospital-related Posts}
\label{sec:identify_hospital_posts}

\subsubsection{Official Discussion}
To identify official discussion from hospitals and hospital systems on
Facebook, we first crawled all hospital homepages from the AHA
database and collected all public-facing Facebook profiles that appeared in
the corresponding Google Knowledge Panel. We supplemented this
collection with a Google search of the hospital name and ``Facebook'',
marking any discovered accounts as associated with a hospital if the
account contained the same metadata as the matching AHA survey
entry. For the 3,845~official Facebook accounts we identified, we
scraped every post created between January~2018 and December~2024,
resulting in 6.9\M~Facebook posts.

\subsubsection{Unofficial Discussion}
To capture patient and provider insights, we studied discussions on the Reddit
social media platform. We collected Reddit data via the
\texttt{Pushshift} project \cite{baumgartner2020pushshift} and started with every
public post and comment on Reddit between January~2018 to December~2024,
covering a total of 1.7\B~posts across roughly
17\M~subcommunities.


To ensure a fair comparison between Reddit and Facebook, we additionally
filtered Reddit posts to only those whose body or comments contained names of
  hospitals in the AHA dataset. This step reduced the Reddit data to 10.3\M~posts
from 555\K~subcommunities.

\subsection{Detecting Ransomware-related Discussion}
\label{sec:detect_ransomware_discussion}

We next identified posts and comments specifically related to the ransomware
attacks in our ground truth corpus. For both Facebook and Reddit, our strategies
were nearly identical: we first curated and evaluated a set of high-precision
keywords from ransomware related posts, and then we time-restricted our
search based on the known date of the attacks. For the Reddit data, we added an
additional final classification step using a large-language model (LLM).

\subsubsection{Keyword Filtering}
\label{sec:keywords}

To build a high-precision keyword list, we first manually identified
ransomware-related posts on Facebook and Reddit by investigating hospital
Facebook pages near the attack dates listed in our ground truth corpus. We
conducted a similar search on Reddit, with a healthcare professional manually
inspecting posts and comments from 127 medical-related subreddits. In short, we
collected 102~ransomware-related Facebook posts and 448~Reddit posts to seed our
initial keywords.

To identify keywords from Facebook posts, we began by running TF-IDF on half of
the posts (51) and grouped keywords into two broad categories: one specifically
for ransomware attacks (e.g., ``ransomware,'' ``cyber attack'') and a broader
set of keywords relating to system outages (e.g., ``system down'', ``lost
access''). Reddit posts lack a consistent post structure and often use loose,
under-specified Internet language (e.g., acronyms, slang, etc.), rendering
TF-IDF insufficient for analysis. As such, for Reddit we manually identified
keywords from a random sample of 224~Reddit posts. We note that we explored many
automated methods for keyword extraction, ranging from simple techniques to
state-of-the-art topic modeling, but found \textit{all} methods required manual
effort to curate the final list of keywords. 


To evaluate the precision of each keyword, we constructed test
datasets with the remaining unobserved posts (51 for Facebook, 224 for
Reddit) combined with an equal number of known false positive posts
about system outages unrelated to ransomware attacks.
Appendix~\ref{appendix:sample_keyword} shows sample keywords used for
Reddit and Facebook from each list as well as the precision of the
Reddit keyword groups.
We included all keywords with a precision higher than
0.75 in the final filter (the bold keyword sub-themes in
Table~\ref{tab:reddit_keywords} in Appendix~\ref{appendix:sample_keyword}). We excluded the remaining
keywords with lower precision because including them captured no
additional ransomware-related posts (they added no value).

Ultimately, we used 108 Facebook keywords and 45 Reddit keywords,
achieving a precision of 0.78 and 0.99 on Facebook and Reddit,
respectively, while capturing 89\% and 80\% of posts in each test-set.
We provide a full list of keywords and filters in the 
paper artifacts.

\subsubsection{Time Matching}
As a final filter, we restricted our analysis to posts that appeared within a
predefined time range after the incident attack date reported in our ground
truth corpus---90~days for Facebook posts and 150~days for Reddit posts.
We determined these time ranges by identifying when the ratio
of filtered ransomware-related posts dropped significantly as the ranges
expanded. A binomial test indicated significant drops at 60~days (p=0.006) and
120~days (p=0.021) for Facebook and Reddit, and we extended the ranges by
an additional 30~days for tail coverage.
We then used these time ranges as a final filter on
the posts.  For each ransomware attack, we associated only those
posts whose dates fell within the Facebook and Reddit time ranges
with that particular attack.





The time matching filter ensured that keyword-filtered posts
appeared within a defined window following the ransomware attack, ensuring
relevance to the specific ransomware attack rather than a random outage. The
time matching filter reduced the Facebook data from 8,331~posts to 309~posts, and
reduced the Reddit data from 6,442~posts to 3,044~posts.

\subsubsection{Classifying Reddit Data with LLMs}
Even after significant filtering, we still observed a substantial number of
false positives in our retrieved Reddit data. To account for this issue, we used
Gemini 2.0 Flash-Lite as a final classifier to determine whether a post was
related to a hospital's ransomware incident, an unexpected system outage, or
neither (Appendix~\ref{appendix:prompt} has the prompt we used for classification). We evaluated the classifier on the
448~manually-collected positive posts and 500~randomly-selected negative posts.
It correctly identified 91 ransomware-related posts, 28 outage-related posts,
and all 500 unrelated posts with a perfect precision of 1.0, producing no false
positives. Given that our goal was not exhaustive detection of such posts, but
rather high-precision identification, we proceeded with this model.

In total, our final dataset contains 1,319~ransomware-related Reddit
posts.\footnote{We manually verified the relevance of these posts to
ransomware attacks with two independent coders during content
analysis.}  We use these Reddit posts together with the 309 Facebook
posts after time matching for our quantitative analysis in
Section~\ref{sec:description}.

\begin{figure*}[t]
\begin{center}
\includegraphics[width=\textwidth]{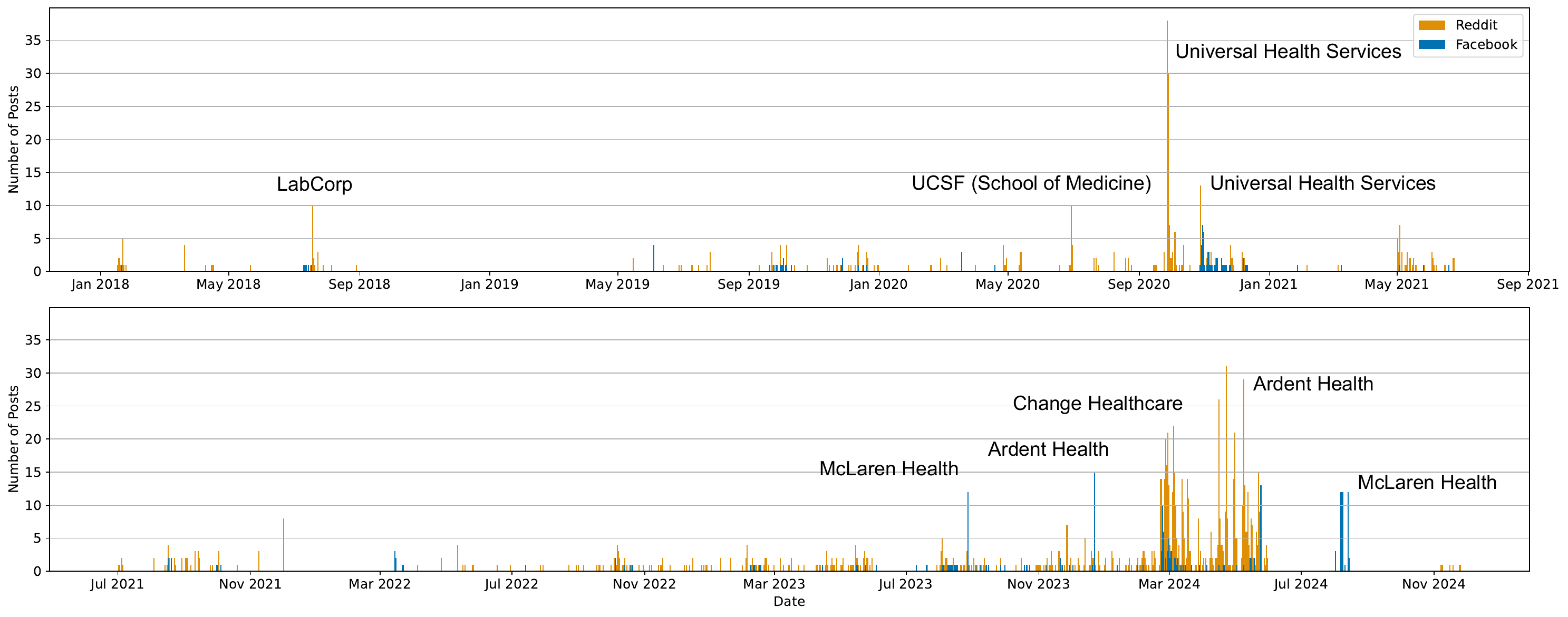}
\end{center}
\caption{\label{fig:longitudinal_figure}Longitudinal frequency of Facebook and
Reddit posts from 2018--2024.
Peaks with more than 10 posts have annotations of the corresponding victim
organizations.}
\end{figure*}

\begin{table}[t]
\begin{center}
\begin{tabular}{lrrrr}
\toprule    
  & \textbf{Facebook} & \textbf{Reddit} \\
\midrule
AHA Hospitals & 6.9\M & 10.3\M \\
Keywords      & 8,331 & 6,442  \\
Time Matching   & 309   & 3,044  \\
LLM           & 309   & 1,319  & \\
Inductive Coding  & 265   & 298    \\
\bottomrule  
\end{tabular}
\end{center}
\caption{\label{tab:data-summary} Summary of filtering steps on posts collected on both platforms.  Inductive coding also identified 123 Facebook and 180 Reddit comments as relevant to their posts.}
\end{table}


\subsection{Content Analysis}
\label{sec:coding}
To examine the \textit{impact} of ransomware attacks on hospitals, providers,
and patients, we built a codebook using inductive coding and thematic analysis
in four stages:


\textbf{Initial Exploration.} One author first familiarized themselves with the
raw text data of all posts by extracting any quotes that explicitly discuss some
downstream effect to the hospital, patients, or providers during an attack.


\textbf{Constructing a Codebook.} We then extracted seven primary themes from
the quotes: computer/network/IT, pharmacy, healthcare facing service, patient
facing service, patient diversion, data breach, and mental health, as well as
subthemes for pharmacy, healthcare facing services, and patient facing services
(full codebook in Appendix~\ref{appendix:codebook}).




\textbf{Coding Posts.}
Two independent coders then applied the codebook to all posts from Facebook and
Reddit, and posts could be tagged with multiple categories.
We excluded all posts that lacked any discussion on impact (64\% of
posts). Our initial
Kupper-Hafner inter-rater reliability score was 0.69, showing the strong
association between the two raters' labels. The raters then resolved any
differences via discussion. Our final dataset for qualitative analysis contains
265~Facebook posts and 298~Reddit posts.


\textbf{Coding Comments.}
Random sampling indicated that the comments to the Facebook and Reddit posts could also have relevant content. Therefore, we also extracted 778~Facebook
comments from 265~Facebook posts and 5,056~Reddit comments from 298~labeled
Reddit posts, and repeated the process of applying the codebook and resolving
conflicts. We excluded comments not related to the impact on healthcare from
labeling, ultimately identifying 123~Facebook and 180~Reddit comments with relevant content.

In the end, after filtering out posts and comments that did not discuss the
impact of each attack, we labeled 265~Facebook posts, 123~Facebook comments,
298~Reddit posts, and 180~Reddit comments.  We use these posts and
comments for our qualitative analysis in Section~\ref{sec:impact}.


\subsection{Attack Target Labeling}
Finally, we categorized each attack target into three types of organization:
``individual hospital'', ``hospital system'', and ``third party''. We distinguished the first two
categories using AHA labels associated with the organization mentioned
in the post. The ``third party'' category included ransomware attack instances
on healthcare companies outside the hospital domain such as Change
Healthcare, which we discuss in more detail in Section~\ref{sec:change}.


%% file: analysis.tex
\section{What can social media tell us about ransomware attacks on hospitals?}
\label{sec:description}

In this section, we analyze our social media data to study the impact of ransomware attacks on hospital systems. We discuss the volume of ransomware attacks covered by our data, quantitatively analyze the discussion surrounding each attack, and detail fundamental differences in how Facebook and Reddit data inform analyses of ransomware attacks.
For this analysis, we use the 309 Facebook and 1,319 Reddit posts that
match our ransomware keywords, timing, and classifier filters
described in Section~\ref{sec:detect_ransomware_discussion}.
Figure~\ref{fig:longitudinal_figure} shows the longitudinal
timing patterns of these posts from 2018 to 2024.

\begin{table}[t]
\small
\begin{center}
  \begin{tabularx}{\columnwidth}{@{}r@{\hspace*{7pt}} r@{\hspace*{2pt}}l@{\hspace*{6pt}} r@{\hspace*{2pt}}l@{\hspace*{6pt}} r@{\hspace*{2pt}}l@{\hspace*{6pt}} r@{\hspace*{2pt}}l@{}}
  \toprule
     \multicolumn{3}{r}{\textbf{Individual}} & \multicolumn{2}{c}{\textbf{System}} & \multicolumn{2}{c}{\textbf{Third}} & \multicolumn{2}{c}{\textbf{Total}} \\
     \midrule
     Facebook Only & 11\% & (24) & 3.3\% & (7) & 0.47\% & (1) & 15\% & (32)\\
     Reddit Only & 25\% & (53) & 23\% & (49) & 25\% & (54) & 74\% & (156)\\
     Both & 5.2\% & (11) & 5.7\% & (12) & 0.47\% & (1) & 11\% & (24)\\
     \midrule
     Total & 42\% & (88) & 32\% & (68) & 26\% & (56) & 100\% & (212) \\
     \bottomrule
\end{tabularx}
\caption{\label{tab:basic_stat_attack} Summary of attacks discussed on each social media platform and victim organization type: individual hospitals, hospital systems, and third-party healthcare companies. The first column reports percentages relative to the total of 212 attacks, and the second reports the corresponding absolute number of attacks.}
\end{center}
\end{table}

\subsection{Attack coverage}

These posts correspond to 212~attacks out of a potential 938~attacks
in the ground truth dataset.  Table~\ref{tab:basic_stat_attack} shows
the distributions of these attack discussions on Facebook and Reddit
for the three different kinds of target organizations.
Social media discussion is spread fairly evenly:
42\%, 32\%, and 26\% of the discussions account
for individual hospitals, hospital systems, and third-party attacks,
respectively. Of the 212~attacks, only 56 (26\%) are discussed on Facebook,
highlighting that official communication from hospitals is much sparser in our
dataset compared to decentralized conversation on Reddit (85\% of attacks appear
on Reddit). Notably, of the 144 hospitals with public Facebook profiles that
were attacked, only 56 (39\%) use Facebook as a mechanism to keep patients
apprised of attack progress.

Since we see posts from only a subset of hospitals that were attacked,
we check whether there are biases in the types of hospitals that post
about attacks.  For each platform we compare the distribution of
organization types for hospitals with social media posts with the
distribution for our ground truth ransomware dataset.  We
randomly selected 50 of the 938 hospitals from the ground truth
dataset as a reference.  Hospitals that post to Facebook show a
significantly different distribution ($\chi^2$=8.1, p=0.018) of
organization types, overrepresenting individual hospitals and
underrepresenting third-party organizations without a social media
presence.  Correspondingly, individual hospitals and clinics are twice
as likely to communicate about attacks via Facebook compared to
hospital systems (16\% vs. 9.0\%). However, Reddit shows no
statistically significant difference in the distribution of individual
hospitals, hospital systems, and third-parties from the randomly
selected reference distribution ($\chi^2$=4.3, p=0.11).

In short, while Reddit generally captures more
attacks than Facebook for all three types of organizations, it
also represents the distribution of attacked organization more accurately.
Facebook discussions are also much less likely to include ransomware attacks on
third-party healthcare providers. Across all attacks, only two third-party
attacks are discussed on Facebook: Change Healthcare and One Blood. In contrast,
55~third-party attacks are discussed on Reddit, including Change Healthcare.
This difference is likely because Facebook discussions from hospitals focus on
attacks on their own infrastructure, even when third-party attacks may be
impacting their ability to provide care. We explore the prominent third-party
attack against Change Healthcare as a case study in
Section~\ref{sec:change}.

\subsection{Initial Discussion Timing}
\label{sec:initial_post}



\begin{figure*}[t]
    \centering
    \subfloat[\centering Facebook]{{\includegraphics[width=0.5\textwidth]{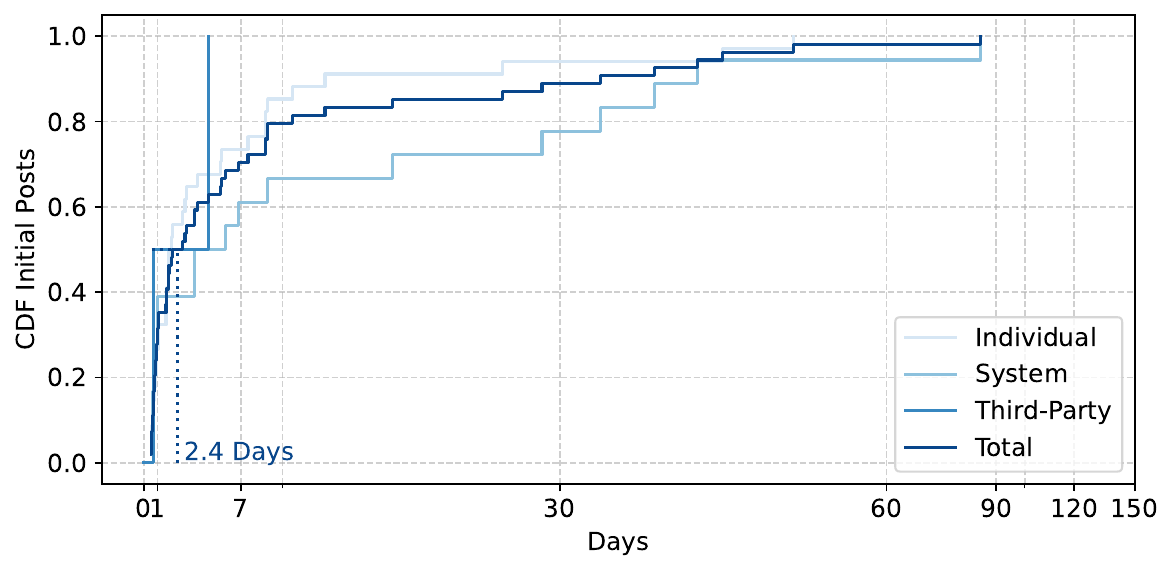} }}%
    \subfloat[\centering Reddit]{{\includegraphics[width=0.5\textwidth]{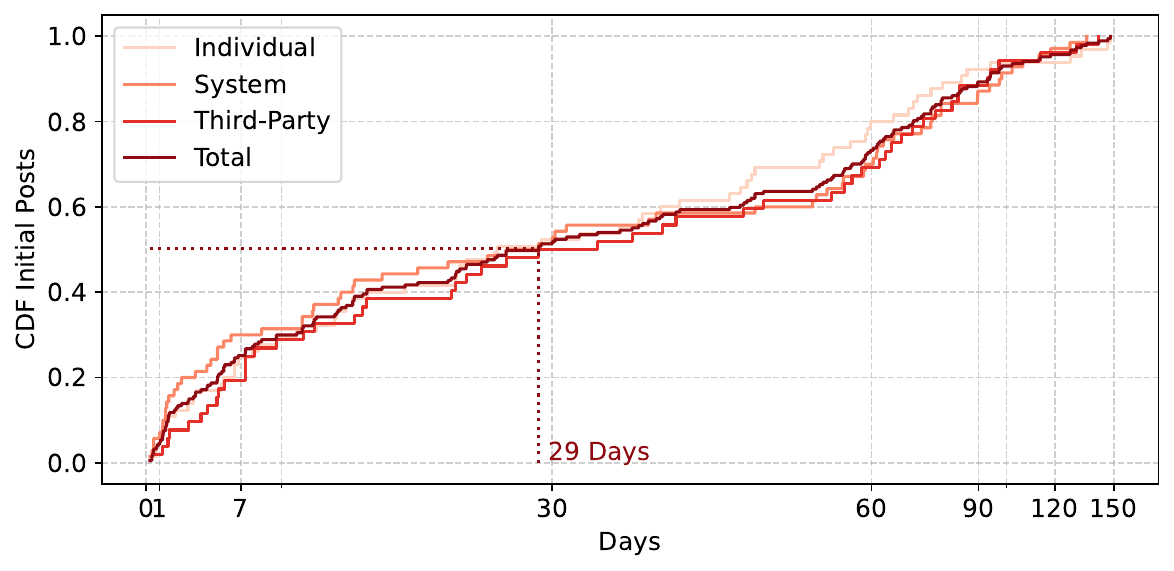} }}%
    \caption{CDFs of the time of initial attack-related posts after the attack date, with different curves for each organization type. While Facebook shows clear differences among the types, Reddit's distributions are more uniform.}%
    \label{fig:cdf_org_type}%
\end{figure*}



We next examine when the initial posts discussing attacks appear online
relative to the known date of attacks. Facebook has a
significantly faster initial post time than Reddit with an average of
2.4~days compared to 29 days (Mann-Whitney U = 8100, p$<$0.005). Most posts on
Facebook occur on the day of the attack (48\%), with a statistically significant
drop in reporting tendency thereafter (one-sided exact Poisson test, p=0.0052).
In contrast, conversation on Reddit is delayed: only 5.0\% of attacks are
discussed within 24~hours. However, Reddit has a large spread of
posts over time per attack.  Figure~\ref{fig:cdf_org_type} shows the
distributions of discussion for each attack target over time. On Facebook, 80\%
of initial posts about an attack appear within 9~days after the attack,
whereas 80\% of post-volume takes approximately 60~days on Reddit.



We also categorize the framing of initial posts
as either ``system outage'' or ``ransomware'',
where posts labeled ``system outage'' report outages without
mentioning the attack. Facebook is more likely to have initial posts as a
``system outage'': 41\% of initial posts on Facebook are characterized as a
``system outage'', with 25\% attack targets never publicly disclosing via
Facebook that they were the victim of a ransomware attack. Reddit, in contrast,
has only 4 (2.2\%) initial posts labeled ``system outage.''

\subsection{Discussion of Attacks}

\subsubsection{Number of Posts Per Attack}
After analyzing the initial posts, we next compare all related posts to identify
broader patterns in ransomware discussions. Table~\ref{tab:all_post_count} shows
the total number of ransomware-related posts made per organization type and the
average number of posts per attack. Attacks discussed on Facebook have an
average of 5.5~posts made about them, compared to 7.3~on Reddit. 

In particular, the average number of posts per attack
for hospital systems is higher than individual hospitals, reflecting
the larger scale of these organizations. Similarly, the average
number of posts for third-party healthcare companies is even higher---an average of 40~posts for the Change Healthcare attack and
12~posts for the other third-party companies---as a single
attack on them can impact many hospitals.  Reddit does not exhibit
this trend, except for the prominent attack on Change Healthcare
which generated 442 related posts.

\subsubsection{Time Length of Posts Per Attack}
\label{sec:conversation-length}

We next study the duration of discussions on each platform.  For each
attack, we use the time from the first post to the last follow-up post
as the discussion duration.  On Facebook, attack discussions last 7.7
days on average, well below the average recovery time for hospitals
affected by ransomware of 17--24 days~\cite{downtimeavg-comparitech,
  gates2024cyber}, suggesting that public conversation on Facebook
about the attack ceases well before attacks are resolved in
practice. Reddit posts for each attack persist significantly longer, 45
days on average.



One explanation for the extended duration of Reddit posts is due to the modality
of the platform: much discussion on Reddit is driven by links to external
sources (e.g., other webpages). Indeed, 820 (62\%) Reddit posts start by
referencing a link to an external source without any additional context added;
798 of these posts are from news media~\cite{bernhardclemm_2023_7651047} and
security journals (based on URL domains). As such, prolonged Reddit content
continues as a \textit{downstream effect} of external discussion---and, as such,
eliciting more discussion when news or articles continue to foreground
ransomware attacks against hospital systems.

\begin{table}[t]
\small
\begin{center}
  \begin{tabularx}{\columnwidth}{lRR|RR|RR|RR}
  \toprule
     \textbf{Days} & \multicolumn{2}{c}{\textbf{Individual}} & \multicolumn{2}{c}{\textbf{System}} &
     \multicolumn{2}{c}{\textbf{Third}} & \multicolumn{2}{c}{\textbf{Change}} \\
      & FB & R & FB & R & FB & R & FB & R \\
     \midrule
     $<$ 1 & 28 & 8 & 43 & 25 & 4 & 7 & 2 & 3\\
     $<$ 7 & 28 & 49 & 39 & 211 & 6 & 34 & 17 & 76\\
     $<$ 30 & 22 & 44 & 78 & 185 & 2 & 36 & 16 & 184\\
     $<$ 60 & 2 & 34 & 9 & 61 & 0 & 23 & 3 & 72\\
     $<$ 90 & 0 & 20 & 8 & 39 & 0 & 22 & 2 & 104\\
     $<$ 120 & --- & 5 & --- & 22 & --- & 10 & --- & 3\\
     $<$ 150 & --- & 11 & --- & 20 & --- & 11 & --- & 0\\
     \midrule
     Total & 80 & 171 & 177 & 563 & 12 & 143 & 40 & 442\\
     \midrule
     Average & 2.1 & 2.7 & 9.3 & 9.2 & 12 & 2.65 & 40 & 442\\
     \bottomrule
\end{tabularx}
\end{center}
\caption{\label{tab:all_post_count} Number of posts appearing
on Facebook (FB) and Reddit (R)
for various time
ranges separated by organization types: individual hospitals,
hospital systems, third-party healthcare companies (not including Change
Healthcare), and Change Healthcare. The last row shows the average number of posts per attack.}
\end{table}

\subsection{Takeaways}
Our quantitative results suggest several takeaways when using social media data
to analyze ransomware attacks. While Facebook posts more quickly identify system
outages caused by ransomware attacks, discussion ends quickly
and stops well before recovery. In contrast, Reddit posts persist much
longer, but they are typically downstream of news reports---which often leave
out smaller hospitals. Wider coverage of larger hospital systems and healthcare
companies is much more prevalently discussed on Reddit compared to Facebook.

\section{Understanding the Impact of Ransomware}
\label{sec:impact}

In this section, we qualitatively analyze the content shared on Facebook and
Reddit to understand the thematic impact of ransomware attacks on hospitals,
patients, and providers.
We use the set of posts and comments we manually coded for content
analysis as described in Section~\ref{sec:coding}, and
Table~\ref{tab:content_count_all} shows the
distribution of attack impact categories from our codebook across the
265~Facebook posts, 123~Facebook comments, 298~Reddit posts,
and 180~Reddit comments we coded. Posts with multiple impact categories are
counted multiple times. Throughout this section, we refer to posts and 
comments collectively as \textit{discussions}.

While Facebook posts decreased by 14\% after the final coding,
Reddit posts decreased by 77\%. The characteristics of each platform disucssed
in Section~\ref{sec:description} explain this substantial difference. The 
larger reduction in Reddit posts is primarily because 62\% of Reddit posts 
referenced headlines from external sources that only captures the occurrence 
of ransomware attacks without providing further details about the impacts. In 
contrast, Facebook contained more posts describing the impacts of the attacks, 
as hospitals reported these impacts while avoiding to explicitly announce the 
attacks themselves.


\subsection{Discussion on Facebook vs. Reddit}

\begin{table}[t]
\small
\begin{center}
  \begin{tabularx}{\columnwidth}{lr@{\hspace*{2pt}}l@{\hspace*{6pt}}r@{\hspace*{2pt}}l@{\hspace*{6pt}}r@{\hspace*{2pt}}l}
  \toprule
   \textbf{Categories} & \multicolumn{2}{c}{\textbf{FBP}} & \multicolumn{2}{c}{\textbf{FBC}} & \multicolumn{2}{c}{\textbf{R}} \\
   \midrule
   \textbf{Computer/Network} & 78\% & (206) & 4.9\% & (6) & 30\% & (144)\\
   \midrule
   \textbf{Pharmacy} & 14\% & (38) & 8.1\% & (10) & 19\% & (93)\\
   Filling Prescription & 10\% & (27) & 6.5\% & (8) & 15\% & (73)\\
   E-prescription & 6.4\% & (17) & 1.6\% & (2) & 6.1\% & (29)\\
   Available Prescription & 4.1\% & (11) & 0\% & (0) & 0\% & (0)\\
   \midrule
   \textbf{Healthcare Facing} & 40\% & (107) & 32\% & (39) & 31\% & (146)\\
   Electronic Health Rec. & 31\% & (82) & 5.7\% & (7) & 14\% & (67)\\
   \Hquad EPIC/Pyxis/etc. & 25\% & (65) &1.6\% & (2)& 8.3\% & (40)\\
   \Hquad Paper Charting & 6.8\% & (18) & 4.1\% & (5) & 7.5\% & (36)\\
   Lab/Imaging & 8.7\% & (23) & 9.8\% & (12) & 6.3\% & (30)\\
   \Hquad Accessing Result & 0.75\% & (2) & 9.8\% & (12) & 4.2\% & (20)\\
   \Hquad Running New Test & 7.9\% & (21) & 0\% & (0) & 3.1\% & (15)\\
   Treatment & 33\% & (87) & 21\% & (26) & 18\% & (84)\\
   \Hquad Current Treatment & 9.1\% & (24) & 12\% & (15) & 15\% & (70)\\
   \Hquad Scheduled Treatment & 27\% & (72) & 17\% & (21)& 3.6\% & (17) \\
  Paycheck for Provider & 0.75\% & (2) & 4.1\% & (5) & 15\% & (72)\\
   \midrule
   \textbf{Patient Facing} & 47\% & (125) & 14\% & (17) & 23\% & (109)\\
   Billing for Patient & 5.3\% & (14) & 4.9\% & (6) & 19\% & (89)\\
   Website/Portal & 16\% & (43) & 2.4\% & (3) & 1.0\% & (5)\\
   Phone Line & 28\% & (73)& 8.1\% & (10) & 3.3\% & (16)\\
   \midrule
   \textbf{Patient Diversion} & 12\% & (33) & 1.6\% & (2) & 5.9\% & (28)\\
   \midrule
   \textbf{Data Breach} & 1.1\% & (3) & 7.3\% & (9) & 23\% & (109)\\
   \midrule
   \textbf{Mental Health} & 0.38\% & (1) & 0\% & (0) & 11\% & (53)\\ 
   \toprule
\end{tabularx}
\end{center}
\caption{\label{tab:content_count_all}Number of
    Facebook posts (FBP), Facebook comments (FBC), and Reddit posts and
    comments (R) that mention each content category. A post or comment
    can appear in multiple categories. The percentages
    are relative to the total for each
    platform: 265 Facebook posts (FBP), 123 Facebook comments (FBC) and
    478 Reddit posts and comments (R).}
\end{table}

Posts on Facebook primarily focus on notifications and instructions directed at
incoming or existing patients, whereas Reddit discussions contain more anecdotal
content, with users sharing personal experiences and seeking advice. The types
of content shared on Facebook are quite different from Reddit:\footnote{Spearman rank
correlation of 0.65, showing only moderate correlation between the two platforms
despite their topic similarity.} the most common type of post on Facebook
is ``Computer/Network'' (78\%) followed by ``Patient Facing''~(47\%), whereas
``Healthcare Facing'' (31\% and 32\%) is the most common
discussion type in Reddit and Facebook comments.

Unsurprisingly, given Facebook is primarily used as an official, public-facing
utility, Facebook posts are typically related to service unavailability for
direct communication with patients, such as access to electronic health records,
scheduled treatments, patient-facing websites, and patient-facing phone lines
(47\% of posts, excluding overlaps). 

Unlike Facebook posts, Facebook comments are
primarily used as a direct communication portal to the hospitals by
patients.  Since the two channels have distinct
demographics (hospitals vs. patients), we explicitly separate
Facebook comments from their posts in
Table~\ref{tab:content_count_all}.  When comparing them, categories
not directly related to patient care in comments show a more
pronounced difference with Facebook posts. For instance,
``Computer/Network'' (78\% and 4.9\%), ``Website/Portal'' (16\% and
2.4\%), and ``Phone Line'' (28\% and 8.1\%) have large differences
between posts and comments.

In contrast, content on Reddit provides a broader perspective that also includes
patient and provider impacts. Indeed, only 5.3\% of Facebook posts were related
to patients or providers compared to 19\% of Reddit discussions. Reddit data tends to
also include more specific examples of impacts like \textit{``Zepbound Savings
Cards aren't working.''}
Interestingly, Reddit also provides more discussion about the downstream impacts
of ransomware attacks, such as data breaches (\textit{``I got a letter from the
hospital regarding a data breach incident''}). Only three posts on Facebook
referenced a data breach compared to 109~such posts on Reddit. In short, similar
to our results in Section~\ref{sec:description}, Facebook and Reddit offer
complementary views on the impact of ransomware attacks.

Since 30\% of all posts and comments in our dataset
discuss the attack against Change
Healthcare, that one attack has a substantial influence on the
distributions of categories for both Facebook and Reddit.
Appendix~\ref{sec:no_change_discussion} shows the content category distribution
when Change Healthcare is excluded. Because the Change Healthcare attack
impacted insurance billing and payment processing, excluding this incident
drastically reduces the representation of the ``Filling Prescription'',
``Paycheck for Provider'' and ``Billing for Patient'' categories. This reduction
is particularly pronounced in Reddit, with percentages for these categories
decreasing from 15\% $\rightarrow$ 3.6\%, 15\% $\rightarrow$ 2.9\%, and 19\%
$\rightarrow$ 4.6\%, respectively.
\subsection{Characterizing Impact}

Our qualitative coding resulted in seven broad themes that highlight
primary areas of impact of ransomware attacks. We detail each of these
themes below.


\subsubsection{Computer/Network/IT System}
Computer and network downtime is often the first visible impact of a ransomware
attack, resulting both from the direct encryption of critical data and from
containment measures taken by the victim hospitals or healthcare companies. IT-related posts appear more frequently on Facebook (78\% vs. 30\%) because
hospitals explicitly mention system disruptions, whereas Reddit users often
reference them indirectly through other categories. However, when the disruption
is explicitly mentioned, both platforms feature similar types of posts.

\subsubsection{Pharmacy}


Following a ransomware attack, pharmacies experience disruptions or
delays in filling prescriptions for various reasons.  While Facebook
and Reddit have a similar number of posts about these pharmacy issues
(14\% and 19\%), Reddit discussions often include details that Facebook posts
lack.  On Facebook, hospitals notify patients that
\textit{``Pharmacies are still operating with limited capabilities.''}
Similarly, patients share their experience with delayed prescriptions
on Reddit. The types of unavailable medication are never specifically
addressed on Facebook, but are frequently mentioned in Reddit in posts
like \textit{``unable to obtain my disabled husband’s nebulizer asthma
  medication''} and \textit{``denying life saving medications for
  cancer and diabetes.''}

\subsubsection{Healthcare Facing}

Healthcare-facing impacts describe services used by providers or treatments
delivered to patients. Common service impacts include electronic health records
and lab testing. Facebook posts communicate these disruptions
primarily as instructions to patients \textit{``to bring current medications or
printed physician orders to the appointment''} or as a notification that
facilities are \textit{``unable to accommodate outpatients.''}
Facebook comments tend to consist of direct questions about appointment status 
in response to cancellation notifications, such as \textit{``I have a CT scan for
Wednesday, will radiology appointments still be canceled?''}

Reddit further provides more detail into the practical and emotional
consequences of these outages. As hospitals revert from electronic to paper
charts, posts made by providers describe heightened risks of medical errors,
fear of losing their licenses, and extreme workload from slower manual
processes. Nurses on the subreddit \texttt{r/nursing} wrote \textit{``Nurses
will face court cases and potential license loss due to cyberattack-related
medical errors''} and \textit{``I had to do a paper admission which was a real
time killer.''} One patient in the subreddit \texttt{r/Austin} also reported about care delays:
\textit{``I couldn't get IV pain relief and waited an hour in
the ER because staff lacked workflow for this.''}

Reddit further shows how missing lab results from hospitals or third-party
companies affect patient lives in various forms, from delayed start dates for
work (\textit{``ransomware delayed my start date; a delay in second sample
postponed me two weeks, costing about \$2,600''}) to delayed critical treatment
(\textit{``she can't start cancer care without the biopsy results''}). One
Reddit post gives more explanation on how their \textit{``father's grade 4
glioblastoma care was delayed due to the missing scan results from ransomware
attack.''} A similar pattern emerges for canceled or rescheduled appointments:
Facebook posts provide brief notifications, while Reddit discussions offer more
detailed, anecdotal content. For example, one Reddit post specified that
\textit{``Oral Chemo \& Radiation treatment''} was rescheduled due to a
ransomware attack. These Reddit posts demonstrate the ransomware impact on
hospitals and individuals by illustrating the exact medical procedures disrupted
and the corresponding consequences.

Finally, Reddit uniquely captures the financial impact on individual healthcare
workers, such as employees reporting payroll system failures: \textit{``system lost the
ability to send direct deposits or paper checks''} or \textit{``payday passed
and everyone’s check was just copied from previous pay period.''}

\subsubsection{Patient Facing}

Patient-facing services, such as websites, portals, and phone lines, are also
often disrupted during ransomware attacks. Facebook posts much more often
mention these types of impacts (43\%) compared to Reddit discussions (4.4\%). However,
in some notable examples of Reddit discussions, physicians described infrastructure
changes as a result of ransomware attacks---\textit{``personal cell phone
numbers to all the floors''}---indicating that internal phone lines were
affected as well as external lines.

One important area where Reddit data is particularly fruitful is in
understanding the impact to patient-side billing. For example, one patient
described on Reddit how he was overcharged due to a ransomware event: \textit{``The bill was \$2,800, much higher due to a cyberattack earlier this year.''}
Ransomware also disrupted insurance verification, sometimes causing delays or
cancellations of treatment. One patient shared: \textit{``I have a time
sensitive procedure on Monday that could cost in the \$15000--\$20000 if not
covered, forcing me to gamble 5 figures on whether to proceed.''}

\subsubsection{Patient Diversion}

Patient and ambulance diversion \cite{dameff23} is also captured on both Facebook (12\%) and Reddit (5.9\%). However, similar to previous topics, Facebook typically shares this impact as a brief notification. Reddit discussions include more details from providers at nearby facilities managing the overflow of patients: \textit{``all diverted patients are now sent to my hospital. Last night, we started with 65 ED holds and spent the 4 hours triaging ICU patients to admit critical cases.''} Another provider wrote:
\textit{``we're so overloaded from taking diverted cases.''}

\subsubsection{Data Breach}
Another area where Reddit data is much more robust is in the discussion of downstream data breach impacts as a result of ransomware (23\% vs. 1.1\% of Facebook posts). Reddit users discussing these impacts frequently offer responses and help-seeking behavior, for example: \textit{``Data Breach - Join Class Actions.''} Some authors in Reddit shared their breach notification letter to confirm validity and seek help: \textit{``Anyone get a Letter from the [hospital name]?''} In contrast, only one hospital, Northfield Hospital \& Clinics, confirmed a data breach within its own institution: \textit{``Patients’ health and financial data may have been compromised.''} Just two other posts came from Hancock County Health System, reporting on a data breach caused by an attack on a third-party organization. Despite the large concern raised about data breaches, hospitals do not use Facebook to confirm the data breaches as a result of ransomware attacks.

\subsubsection{Mental Health}
Finally, Reddit discussions are uniquely helpful in identifying the distinct mental and emotional strain put on patients and providers during ransomware attacks (11\% vs. 0.38\%~posts on Facebook). Providers face heavy workloads without functional computer systems. One nurse posted on Reddit: \textit{``The systems still down and will be for weeks apparently ... This past stretch of shifts is making me want to leave.''}
Patients experience financial burdens from inaccessible insurance, delayed treatments,
and leaked personal data. Patients also experience psychological stress from delayed treatment or medication, even if they are non-urgent: \textit{``She's missed 2 doses and may miss 2-3 more. Could this harm her multiple sclerosis treatment?''} and \textit{``I was afraid I'd relapse into bad habits without medication.''} Reddit discussions highlight the psychological impact of a ransomware attack lacking in official reports.


\subsection{Impact Over Time}

\begin{figure}
\begin{center}
\includegraphics[width=0.45\textwidth]{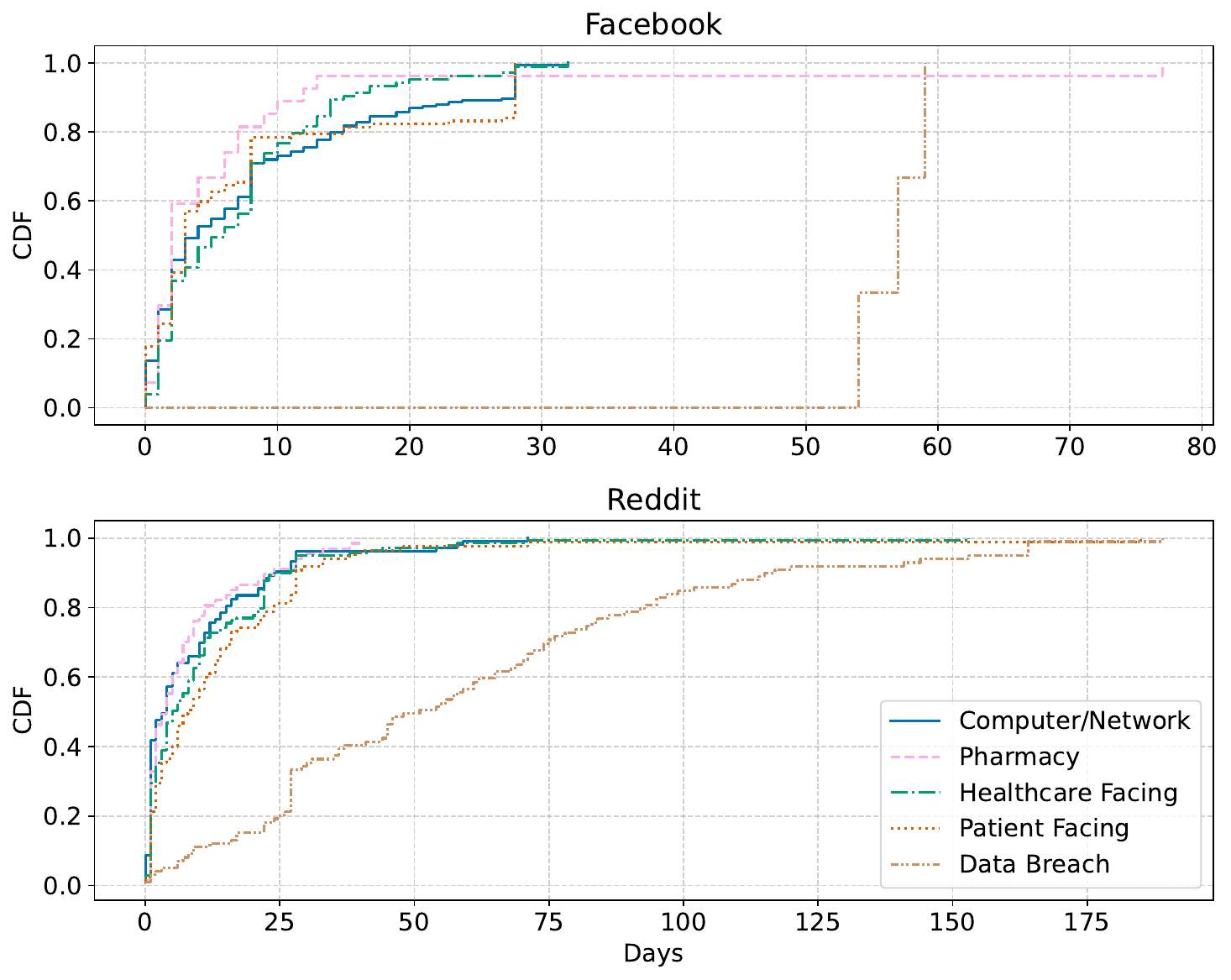}
\end{center}
\caption{\label{fig:content_time_delta} CDFs of the times of posts after the attack date for the five largest categories.}
\end{figure}

We next examined how frequently different content categories appeared relative
to the time after attacks. Figure~\ref{fig:content_time_delta} shows two CDF
graphs for the frequency of categories for Reddit (top) and Facebook (bottom).
Each curve represents five big categories from Table~\ref{tab:content_count_all} that
showed the most prominent differences in distribution.
This time trend in
popular discussion topics helps present the narrative flow in what transpires
during the attack. Both social media platforms show an early appearance in
discussions about healthcare, pharmacy, and patient-facing services immediately
after an attack. However, the frequency of patient-facing services, such as
phone lines or websites, declines more quickly than healthcare or pharmacy
services. Reddit has longer discussions on healthcare and pharmacy outages,
showing that oftentimes outages last longer than apparent from hospital posts on
Facebook. These results are consistent with our earlier finding that Facebook
follow-up posts end sooner than the average recovery period
(Section~\ref{sec:conversation-length}).

Unlike Facebook, where conversation mostly dies down after 30 days, Reddit has a
longer tail of conversation that shifts its focus. Around 27 days after the
attack, the focus of posts on Reddit shifts from service outages to data
breaches.
Also consistent with Section~\ref{sec:conversation-length}, as more news on the
data breaches from attacks comes out, there is a resurgence of discussions on
Reddit corresponding with the shift in the topic of conversation to data
breaches.

\subsection{Takeaways}
This section shows how Facebook posts primarily provided urgent notifications and instructions for patients, reflecting the need of hospitals to maintain safe care while operating at reduced capacity during an attack.
The pattern observed in the time of posting again reflects this result,
where hospitals quickly reported system outages to divert non-emergency patients. However, diversity in Facebook content remained limited, as organizations were reluctant to disclose details about attacks, especially data breaches. Reddit discussions were more anecdotal and detailed, highlighting specific consequences patients experienced during an attack
and the resulting disruptions to their daily lives or medical procedures. 

%% file: case-study.tex
\section{Case Study: Change Healthcare}
\label{sec:change}

The ransomware attack on Change Healthcare in February 2024 was one of the largest attacks against the U.S.\ healthcare system, affecting 192~million individuals and leading to billions of dollars in damages~\cite{change-hipaa-journal, change-ocr-report}. Change Healthcare serves as a clearinghouse for healthcare transactions across the country and operates as a subsidiary of United Healthcare, processing roughly 15\B~transactions a year~\cite{change-aha-survey}. In this section, we apply our social media vantage point and examine how leveraging social media data provides more clarity on the impact of the attack across the entire healthcare system.
For this analysis we use the set of posts that match our keywords
(Section~\ref{sec:detect_ransomware_discussion}) and explicitly
mention Change Healthcare.
This single attack alone accounts for 30\% of these posts: 40~posts (from 32 hospitals) on Facebook and 442~posts on Reddit (across 160 subreddits).


\subsubsection{Online discussion}
\begin{figure}[t]
\begin{center}
\includegraphics[width=0.45\textwidth]{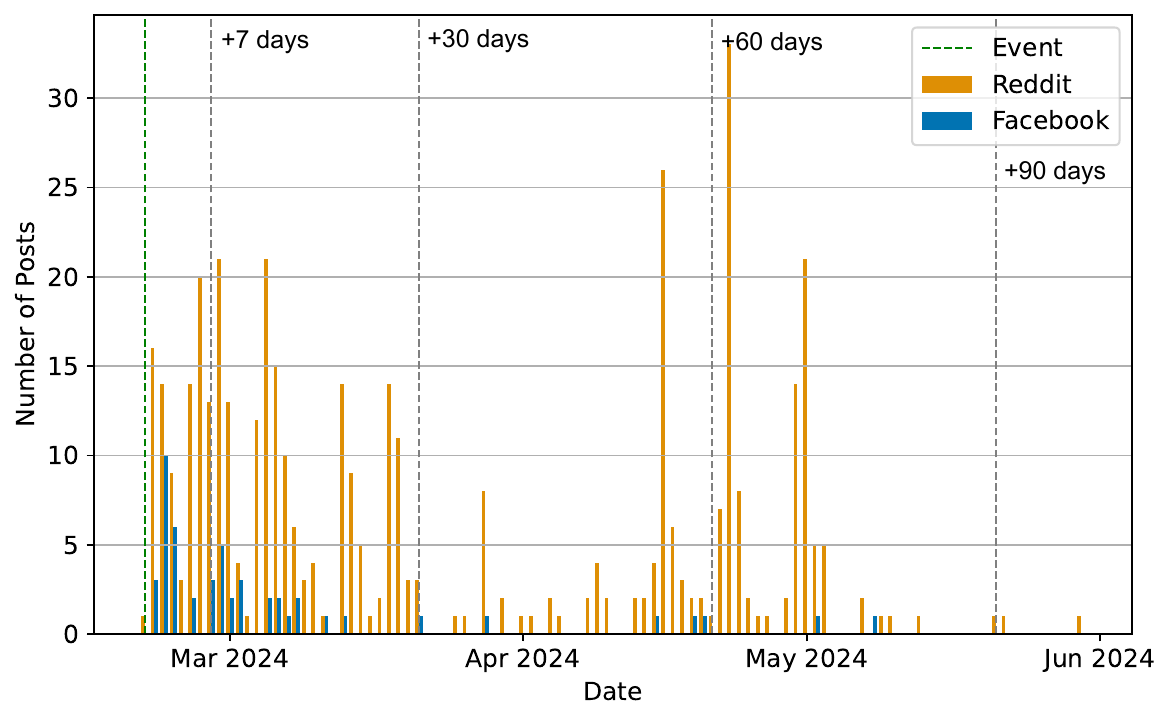}
\end{center}
\caption{\label{fig:onlychange_timeline} Longitudinal frequency of Facebook and Reddit posts related to the Change Healthcare attack. A green dotted line marks the attack date and gray lines mark +N days.}
\end{figure}

In contrast to the other ransomware attacks, discussion patterns about the
Change Healthcare attack are near identical on Facebook and Reddit, and
discussions occur immediately on both platforms (within 24 hours of the attack).
Figure~\ref{fig:onlychange_timeline} shows a timeline of attack discussion for
both Facebook and Reddit in daily granularity. Facebook discussion again tapers
quickly after initial posts by hospitals but
discussion on Reddit about the attack is long-tailed, with another peak
$\sim$60~days after the attack. This peak occurs in reaction to news articles
about Change Healthcare paying the ransom (7.5\% of Reddit posts at the peak).

\subsubsection{Affected hospitals} 
Among the 32 hospitals that report system disruptions, 14 are military and veterans’ healthcare facilities (Naval Medical Center, Veterans Affairs, etc.),
9~are individual hospitals, and 9~are hospital systems. Military healthcare facilities not only report disruptions more frequently but also tend to post earlier notifications:
63\% of the posts made within the first 7~days are from military facilities. While 77\% of Facebook posts on third-party attacks are about Change Healthcare, the 32 hospitals identified via Facebook represent only a small fraction of those actually affected, as 73\% of 1,000 hospitals reported direct operational disruption through the AHA survey~\cite{change-aha-survey}. Reddit discussions do not identify specific hospitals or pharmacies, likely because the Change Healthcare outage caused widespread national outages. However, the discussions revealed its impact on local pharmacies as well as hospitals, effects not captured in formal surveys.




\subsubsection{Impact}
Finally, we shift the analysis to the coded labels
(Section~\ref{sec:coding}) of posts that mention Change Healthcare and
their comments.
The primary impacts of the Change
Healthcare attack were to pharmacy (86\%), patient billing (71\%), and provider
paycheck~(77\%). Patients and providers alike were impacted, notably in
processing often life-saving medication. We observed a significant fraction of
posts (15\%) about patients unable to get essential, life-saving medication
without having to ``pay out of pocket.'' Table~\ref{tab:medication_purpose}
shows a sample of some of the quotes from patients and providers about these
impacts. While much widespread reporting focuses on the financial impact to
hospitals and the healthcare system writ large, social media analyses can also
help to bridge the gap to the real, human impact of cybersecurity incidents.


The impact is not limited to providers and patients, however. Hospitals and
hospital systems were also caught downstream of this attack with little
recourse. Many were unable to file insurance claims for compensation, in some
cases impacting employee salaries or hospital cash flow. \textit{``We have so
many unpaid claims, and insurers say they never received them,''}
\textit{``no one in my company has been paid yet--it's been a full month.''}
Compared to the millions of ransom payments, these Reddit quotes show how
relatively small monetary loss can still be detrimental, especially to smaller
local hospitals or pharmacies. It helps demonstrate how the downtime operation
loss impacts individuals beyond the big numbers.


\begin{table}[t]
\small
\begin{center}
  \begin{tabularx}{\columnwidth}{p{2in}r}
  \toprule
  \textbf{Quotes on Affected Medication} & \textbf{Affected Treatment} \\
  \midrule
  \textit{Someone left trexall and mercaptopurine behind because they cost over \$1,000.} & Chemotherapy \\
  \midrule
    \textit{Went to pick up my insulin today but couldn’t get insurance pricing—would have to pay \$1,000 out of pocket.} & Type-1 Diabetes \\
    \midrule
    \textit{Her Copaxone is cover by insurance, but they couldn't fill it. The only other pharmacy has it at an astronomical cost.} & Multiple Sclerosis \\
  \bottomrule
\end{tabularx}
\end{center}
\caption{\label{tab:medication_purpose} Reddit quotes identifying affected medication and their purpose.}
\end{table}

%% file: discussion.tex
\section{Discussion and Conclusion}
While research on social media has proven fruitful in characterizing
the scope and impact of many real-world events, such as natural
disasters \cite{houston2015social}, political revolutions
\cite{wolfsfeld2013social}, and even novel cybersecurity threats
\cite{mittal2016cybertwitter}, our work is the first to leverage
social data to understand the impact of ransomware attacks in a
healthcare setting. What makes this approach
particularly effective in healthcare is precisely that it surfaces a
broad array of activities and outcomes that are not captured in
traditional reporting mechanisms.  For example, U.S.\ hospitals are
required---under the HIPPA Breach Notification rule---to
disclose data breaches to both the Department of Health and Human
Services and affected individuals, but only concerning its timing,
the nature of the Protected Health Information (PHI) that was
accessed, and a brief description of steps being taken to investigate
and mitigate.  From such reports, it is impossible to gauge the
clinical impact of such episodes, what kinds of patient services
were degraded or unavailable, or the situational understanding of
staff and patients of their circumstances at the time.  Indeed,
there are strong liability and reputational incentives for hospitals
\emph{not} to provide such information voluntarily.  Our dataset
thus is one of the only sources of such unique qualitative
information, opening new possibilities both for characterizing
ransomware impact and, in turn, constructing more targeted
responses to such incidents. In this context, we synthesize the
implications of the findings of our study.

\subsection{Complementary Vantage Points} 
While some attacks were discussed on both Facebook and Reddit, both platforms
together offered complementary vantage points to studying ransomware.
Facebook posts appear earlier and often highlight impact to smaller hospitals
and clinics, while Reddit discussion is often prolonged and offers significantly more
anecdotal information on each attack. Conversely, Reddit discussion is often
downstream of news articles and other sources of reporting, whereas Facebook
posts are official communication from hospitals and offer a first-hand view of
the impact of an attack. Each platform serves different use cases: hospitals use
Facebook primarily to keep patients apprised of hospital status, while patients
and providers flock to Reddit to discuss personal impact on the ground. As such,
no one view can provide a comprehensive look at the impact of a ransomware
attack---both vantage points are necessary to develop a fully realized picture. 

\subsection{Help-Seeking Behaviors on Reddit}
Prior work in the computer security community has studied help-seeking
behaviors through studying Reddit data \cite{wei2024understanding},
highlighting how studying help-seeking behaviors can identify opportunities to
deploy more resources to those in need.  We also observe similar behaviors from
users on Reddit when they are faced with ransomware attacks. Patients and
providers alike turned to Reddit to seek help in their individual
circumstances. For example, during the Change Healthcare attack, patients sought alternative pharmacies to get medications filled and providers looked
for ways to get insurance claims filed and paid for. In other
cases, providers sought both help and validation as they attempted to chart
patients without EMR access. In the face of new attacks, leveraging social data
to examine the pain points for patients and providers can guide
resource allocation to improve patient care or highlight policy gaps that might
need addressing in preparation for future disruptions.

%

\subsection{Limitations}
Our work is not without limitations. For one, while we find significant promise
in leveraging social data, we only covered 22\% of potential attacks in our
ground truth corpus. While the attacks we studied were diverse across many
aspects (e.g., hospital type, size of attack), our methods can only be applied
when discussion about attacks occur more widely. Our data vantage point is also
potentially limited---if hospitals posted about ransomware attacks and
eventually deleted the posts, they would not appear in our corpus. Similarly,
if posts were removed from Reddit or Pushshift, these would also not appear in
our corpus. Our work also focused exclusively on ransomware attacks against
U.S.-based hospitals and discussion in English---more work is required to examine
how ransomware affects hospitals around the world. Finally, one critical
limitation is that the veracity of Reddit data is challenging to verify:
while we could corroborate some stories of hospital impact (e.g., through
independent news articles), we primarily \textit{trust} that the stories posted
to Reddit are true and not fabrications. This limitation is fundamental to
all research that leverages social data.

\subsection{Ethical Considerations}
We carefully evaluated the ethical implications of our methodological choices
in the course of conducting this work. In particular, we sought to minimize
potential risks of leveraging social data to study the impact of ransomware
attacks. The primary stakeholders we considered were the targeted hospitals,
the patients and providers served by those hospitals, and the social media
services (Facebook and Reddit) that host the posts we analyzed.

\subsubsection{Hospitals and Health Systems}
Our analysis focuses on well-documented ransomware attacks against hospitals
and health systems. We stress that all Facebook accounts we studied are
official organization accounts and their posts reflect each hospital's official
communication policy (i.e., not about individuals). For clarity and to
appropriately demonstrate the impact of these attacks, we chose to include the
names of these targets in our paper, as these attacks are public knowledge and
additional risk of attention or exposure would be minimal. The hospital metadata
comes from a purchased AHA dataset, however, and therefore cannot be released.

\subsubsection{Patients and Providers}
In our work, we analyzed hundreds of posts submitted to Reddit, many of which
included personal anecdotes or stories from patients and providers impacted by
ransomware attacks. To protect the privacy of posters, we primarily 
report aggregate results in this paper. In cases where we do share individual
snippets from Reddit posts to highlight a result, we are careful to exclude
any content that a reasonable person might find injurious to the poster (e.g.,
embarrassing, libelous, etc.). In addition, we do not maintain any identifying
information about the posters in our analysis and do not report usernames or
other potentially deanonymizing indicators in our work. To reduce the risk 
of reidentification of the poster, we paraphrased all quotes included in the paper.

\subsubsection{Facebook and Reddit}
Our work fundamentally relies on data sourced from Facebook and Reddit. All
data we used in this paper was public and pre-existing, and we did not solicit
any private information to conduct our analysis. The datasets we collected from
Facebook used an approved API. 


%% file: acks.tex
\section*{Acknowledgments}

We thank our anonymous reviewers for their insightful suggestions and
feedback.  Many thanks also to Cindy Moore and Jennifer Folkestad for
operational and administrative support of our research.  Funding
was provided by the Advanced Research Projects Agency for
Health (ARPA-H) grant SP4701-23-C-0075, the Irwin Mark and Joan Klein
Jacobs Chair in Information and Computer Science, the CSE
Professorship in Internet Privacy and/or Internet Data Security, the
Sheldon Furst Chair in Anesthesiology, and operational support from
the UCSD Center for Networked Systems.

%% file: checklist.tex
\newcommand{\answerYes}[1]{\textcolor{blue}{#1}} 
\newcommand{\answerNo}[1]{\textcolor{teal}{#1}} 
\newcommand{\answerNA}[1]{\textcolor{gray}{#1}} 
\newcommand{\answerTODO}[1]{\textcolor{red}{#1}} 

\section*{Paper Checklist}

\begin{enumerate}

\item For most authors...
\begin{enumerate}
    \item  Would answering this research question advance science without violating social contracts, such as violating privacy norms, perpetuating unfair profiling, exacerbating the socio-economic divide, or implying disrespect to societies or cultures?
    \answerYes{Yes}
  \item Do your main claims in the abstract and introduction accurately reflect the paper's contributions and scope?
    \answerYes{Yes}
   \item Do you clarify how the proposed methodological approach is appropriate for the claims made? 
    \answerYes{Yes}
   \item Do you clarify what are possible artifacts in the data used, given population-specific distributions?
    \answerYes{Yes, we identify the sources of data in our method section.}
  \item Did you describe the limitations of your work?
    \answerYes{Yes, we discuss limitation in coverage and verification of social media data.}
  \item Did you discuss any potential negative societal impacts of your work?
    \answerYes{Yes, see ``Ethical Considerations''.}
      \item Did you discuss any potential misuse of your work?
    \answerNA{NA}
    \item Did you describe steps taken to prevent or mitigate potential negative outcomes of the research, such as data and model documentation, data anonymization, responsible release, access control, and the reproducibility of findings?
    \answerYes{Yes, we describe our methods in ethical considerations section.}
  \item Have you read the ethics review guidelines and ensured that your paper conforms to them?
    \answerYes{Yes}
\end{enumerate}

\item Additionally, if your study involves hypotheses testing...
\begin{enumerate}
  \item Did you clearly state the assumptions underlying all theoretical results?
    \answerNA{NA}
  \item Have you provided justifications for all theoretical results?
    \answerNA{NA}
  \item Did you discuss competing hypotheses or theories that might challenge or complement your theoretical results?
    \answerNA{NA}
  \item Have you considered alternative mechanisms or explanations that might account for the same outcomes observed in your study?
    \answerNA{NA}
  \item Did you address potential biases or limitations in your theoretical framework?
    \answerNA{NA}
  \item Have you related your theoretical results to the existing literature in social science?
    \answerNA{NA}
  \item Did you discuss the implications of your theoretical results for policy, practice, or further research in the social science domain?
    \answerNA{NA}
\end{enumerate}

\item Additionally, if you are including theoretical proofs...
\begin{enumerate}
  \item Did you state the full set of assumptions of all theoretical results?
    \answerNA{NA}
	\item Did you include complete proofs of all theoretical results?
    \answerNA{NA}
\end{enumerate}

\item Additionally, if you ran machine learning experiments...
\begin{enumerate}
  \item Did you include the code, data, and instructions needed to reproduce the main experimental results (either in the supplemental material or as a URL)?
    \answerYes{Yes, we included the LLM prompt used in the Appendix.}
  \item Did you specify all the training details (e.g., data splits, hyperparameters, how they were chosen)?
    \answerYes{Yes, see ``Classifying Reddit Data with LLMs'' in Section 3.3.}
     \item Did you report error bars (e.g., with respect to the random seed after running experiments multiple times)?
    \answerYes{Yes, see ``Classifying Reddit Data with LLMs'' in Section 3.3.}
	\item Did you include the total amount of compute and the type of resources used (e.g., type of GPUs, internal cluster, or cloud provider)?
    \answerYes{Yes, see ``Classifying Reddit Data with LLMs'' in Section 3.3.}
     \item Do you justify how the proposed evaluation is sufficient and appropriate to the claims made? 
    \answerYes{Yes, see ``Classifying Reddit Data with LLMs'' in Section 3.3.}
     \item Do you discuss what is ``the cost`` of misclassification and fault (in)tolerance?
    \answerYes{Yes, see ``Classifying Reddit Data with LLMs'' in Section 3.3.}
  
\end{enumerate}

\item Additionally, if you are using existing assets (e.g., code, data, models) or curating/releasing new assets, \textbf{without compromising anonymity}...
\begin{enumerate}
  \item If your work uses existing assets, did you cite the creators?
    \answerYes{Yes, we identify the sources of data in our method section.}
  \item Did you mention the license of the assets?
    \answerNA{NA}
  \item Did you include any new assets in the supplemental material or as a URL?
    \answerNA{NA}
  \item Did you discuss whether and how consent was obtained from people whose data you're using/curating?
    \answerYes{Yes, see ``Ethical Considerations'' in Section 7.4.}
  \item Did you discuss whether the data you are using/curating contains personally identifiable information or offensive content?
    \answerYes{Yes, we discuss about anonymizing the data in ``Ethical Considerations.''}
\item If you are curating or releasing new datasets, did you discuss how you intend to make your datasets FAIR (see \cite{fair})?
\answerNA{NA}
\item If you are curating or releasing new datasets, did you create a Datasheet for the Dataset (see \cite{gebru2021datasheets})? 
\answerNA{NA}
\end{enumerate}

\item Additionally, if you used crowdsourcing or conducted research with human subjects, \textbf{without compromising anonymity}...
\begin{enumerate}
  \item Did you include the full text of instructions given to participants and screenshots?
    \answerNA{NA}
  \item Did you describe any potential participant risks, with mentions of Institutional Review Board (IRB) approvals?
    \answerNA{NA}
  \item Did you include the estimated hourly wage paid to participants and the total amount spent on participant compensation?
    \answerNA{NA}
   \item Did you discuss how data is stored, shared, and deidentified?
   \answerNA{NA}
\end{enumerate}

\end{enumerate}

%% file: real_appendix.tex
\section{Keyword Filtering}
\label{appendix:sample_keyword}%

Section~\ref{sec:keywords} describes our overall methodology for
identifying and using keywords to filter social media posts to those
related to ransomware attacks on healthcare organizations.  In this
section of the appendix we include two tables that provide more
context and detail on these keywords.
Table~\ref{tab:example_keywords_specific} lists example keywords used
to detect ransomware-related discussions in Facebook and Reddit posts,
and Table~\ref{tab:reddit_keywords} shows the precision scores for the
Reddit keywords we evaluated.

\begin{table}[h]
    \centering
    \subfloat[Facebook]{
    \begin{tabularx}{\columnwidth}{YY}
  \toprule
    \textbf{Ransomware Theme} & \textbf{Outage Theme} \\
    \midrule
	cyberattack & information technology\\
     cyber attack & be down\\
     ransomware & computer system\\
     cybersecurity incident & postpone \\
     threat actor & phone service \\
     \bottomrule
\end{tabularx}
    }\\[1em]
    \subfloat[Reddit]{\begin{tabularx}{\columnwidth}{YY}
  \toprule 
  \textbf{Ransomware Theme} & \textbf{Outage Theme}\\
    \midrule
	ransomware & goes down\\
  cyberattack & technical issue\\
  cybersecurity incident & written order\\
  data breach & cannot login\\
    cybersecurity matter & network outage \\
     \bottomrule
\end{tabularx}}%
    \caption{\label{tab:example_keywords_specific}Example ransomware-themed and outage-themed keywords used for
    Facebook and Reddit.}%
\end{table}



\begin{table}[h]
\begin{center}
  \begin{tabular}{ l l r }
  \toprule
    \textbf{Theme} & \textbf{Sub-Theme} & \textbf{Precision} \\
    \midrule
    Ransomware & \textbf{\texttt{ransomware}} & 1.0 \\
    \midrule
    \multirow{11}{*}{System Outage} &\textbf{\texttt{TECH issue}} & 1.0 \\
    & \textbf{\texttt{outage}} & 1.0 \\
    & \textbf{\texttt{paper}} & 1.0 \\
    & \textbf{\texttt{sign in}} & 0.86 \\
    & \textbf{\texttt{NOUN/GO down}} & 0.83 \\
    & \texttt{BE down} & 0.42 \\
    & \texttt{access} & 0.40 \\
    & \texttt{BE out} & 0.26 \\
    & \texttt{delay} & 0.21 \\
    & \texttt{BE issue} & 0.13 \\
    & \texttt{work} & 0.05 \\
    \bottomrule
\end{tabular}
\caption{\label{tab:reddit_keywords} Precision scores for all sub-themes of Reddit keywords with scores higher than 0.75. Bolded sub-themes are used for final keywords.}
\end{center}
\end{table}

\section{Identifying Hospital Names from AHA Data}
We first identified common suffixes for hospital names like ``hospital'',
``health'', ``medical'', and ``clinic''. We then generated a hospital name
filter using three rules:

\begin{itemize}
   \item If the hospital has an unique name, we remove the common titles from the keyword (e.g., $\texttt{A Hospital} \rightarrow \texttt{A}$).
   \item If the hospital has a non-unique name, like a location or a verb, we use all common title keywords ($\texttt{B Hospital} \rightarrow \texttt{B Hospital, B Health,}\cdots$).
   \item If two different institutions exist with a similar name, like \texttt{X Hospital} and \texttt{X Healthcare}, we use the full name as a keyword.
\end{itemize}

This filter prevents false negatives from trivial mistakes of interchanging common words like ``health'' or ``healthcare'', and prevents common verb names like ``prospect'' or ``ardent'' from producing false positives.

\section{LLM Prompt}
\label{appendix:prompt}

You are a three-way classifier determines if the following Reddit post is related to hospital ransomware attack or unexpected system down based on the Reddit post title and content.

\begin{enumerate}
  \item The post is related if it talks about the hospital getting hit by ransomware attack and having system down
  \item The post is related if it talks about the hospital has an unexpected system down
\end{enumerate}

Reddit Post Title: ``\{\texttt{title}\}''. 

Reddit Post Content: ``\{\texttt{description}\}''. 

Is this Reddit post related to hospital ransomware attack or hospital unexpected system down or neither?

\section{Content Analysis Codebook}
\label{appendix:codebook}

The code applied to all posts in Facebook and Reddit dataset.
\begin{enumerate}
	\setlength{\itemsep}{1pt}
  	\setlength{\parskip}{1pt}
	\item \textbf{Computer/Network}: Computer, network, or IT system downtime is mentioned explicitly.
	\item Pharmacy
	\begin {itemize}
		\item \textbf{Filling Prescription}: Any disruption in filling or obtaining prescriptions, including delays.
		\item \textbf{E-Prescription}: Disruptions in prescription processing (sending, receiving, or filing), including e-prescription downtime or reference to paper prescription.
		\item \textbf{Available Prescription}: Information on the availability of prescriptions, such as restrictions to ``urgent'' medications.
	\end{itemize}
	\item Healthcare Facing-Electronic Health Records
	\begin{itemize}
		\item \textbf{EPIC/Pyxis/etc.}: Disruptions on accessing electronic medical health records.
		\item \textbf{Paper Charting}: Any disruption explicitly mentioning the use of manual or paper charting as a result of ransomware.
	\end{itemize}
	\item Healthcare Facing-Lab/Screening/Imaging
	\begin{itemize}
		\item \textbf{Accessing Result}: Disruptions in accessing labortabory results.
		\item \textbf{Running New Test}: Inability to conduct new test or labortabory procedures due to ransomware.
	\end{itemize}
	\item Healthcare Facing-Treatment/Appointment
	\begin{itemize}
		\item \textbf{Current Treatment}: Disruptions in patient treatment caused by ransomware, including delays (e.g., extended wait times) during hospital visits or explicitly mentioned indirect impacts resulting from increased provider workload.
		\item \textbf{Scheduled Treatment}: Disruptions affecting scheduled appointments, including rescheduling, cancellations, or delays.
	\end{itemize}
	\item Healthcare Facing-\textbf{Paycheck for Provider}: Disruptions that results in financial impacts on healthcare workers or hospitals, excluding ransom payments.
	\item Patient Facing
	\begin {itemize}
		\item \textbf{Billing for Patient}: Disruptions that result in financial impacts on patients, excluding scams related to breached information. This includes issues such as being charged incorrect amounts or being unable to afford treatment or medication.
		\item \textbf{Phone Line}: Disruptions on internal or external phone lines.
		\item \textbf{Website/MyChart/Portal}: Disruptions on web-based services that are reachable from patients.
	\end{itemize}
	\item \textbf{Patient Diversion}: Any kind of patients or ambulance diversions, including contents related to ``spillover effects'' from near-by hospitals.
	\item \textbf{Data Breach}: Any contents regarding data breaches from ransomware attacks. It includes expression of concern from patients.
	\item \textbf{Mental Health}: Any contents mentioning authors' pyschological stress.
\end{enumerate}





\section{Discussion on Facebook vs. Reddit without Change Healthcare Attack}
\label{sec:no_change_discussion}

We present Table~\ref{tab:content_count_all_no_change} that shows the
distribution of attack impact categories like Table~\ref{tab:content_count_all}.
However, we excluded the Facebook/Reddit posts and comments related to Change
Healthcare attack that occurred in February 2024.

The Change Healthcare attack disrupted the insurance verification and filing
process, most significantly impacting ``Paycheck for Provider'' and ``Billing
for Patient'' categories the most. Insurance reimbursements to providers were
suspended, straining provider cash flow, while patients were unable to have
their copayments or insurance verified and were required to pay out-of-pocket.
This financial disruption further affected the ``Filling Prescription'' and
``Current Treatment'' categories, as patients suddenly could not afford their
medications or ongoing treatments. The impact of Change Healthcare attack is
further describe in Section~\ref{sec:change}. We observe the resulting reduction
in these categories in the in Table~\ref{tab:content_count_all_no_change}
relative to Table~\ref{tab:content_count_all}.

\begin{table}[h]
\small
\begin{center}
  \begin{tabularx}{\columnwidth}{lr@{\hspace*{2pt}}l@{\hspace*{6pt}}r@{\hspace*{2pt}}l@{\hspace*{6pt}}r@{\hspace*{2pt}}l}
  \toprule
   \textbf{Categories} & \multicolumn{2}{c}{\textbf{FBP}} & \multicolumn{2}{c}{\textbf{FBC}} & \multicolumn{2}{c}{\textbf{R}} \\
   \midrule
   \textbf{Computer/Network} & 79\% & (179) & 4.9\% & (6) & 23\% & (110)\\
   \midrule
   \textbf{Pharmacy} & 4.9\% & (11) & 3.5\% & (4) & 5.9\% & (28)\\
   Filling Prescription & 3.5\% & (8) & 1.8\% & (2) & 3.6\% & (17)\\
   E-prescription & 1.3\% & (3) & 1.8\% & (2)& 2.9\% & (14)\\
   Available Prescription & 2.2\% & (5) & 0\% & (0) & 0\% & (0)\\
   \midrule
   \textbf{Healthcare Facing} & 47\% & (107) &35\% & (39) & 24\% & (116)\\
   Electronic Health Rec. & 36\% & (82) & 6.2\% & (7) & 13\% & (62)\\
   \Hquad EPIC/Pyxis/etc. & 29\% & (65) &1.8\% & (2)& 7.5\% & (36)\\
   \Hquad Paper Charting & 8.0\% & (18) & 4.4\% & (5) & 7.1\% & (34)\\
   Lab/Imaging & 10\% & (23) & 11\% & (12) & 6.1\% & (29)\\
   \Hquad Accessing Result & 0.88\% & (2) & 11\% & (12) & 4.2\% & (20)\\
   \Hquad Running New Test & 9.3\% & (21) & 0\% & (0)& 2.9\% & (14)\\
   Treatment & 38\% & (87) & 23\% & (26)& 13\% & (60)\\
   \Hquad Current Treatment & 11\% & (24)& 13\% & (15) & 9.6\% & (46)\\
   \Hquad Scheduled Treatment & 32\% & (72)& 19\% & (21)& 3.6\% & (17) \\
   Paycheck for Provider & 0.44\% & (1) & 44\% & (5)& 2.9\% & (14)\\
   \midrule
   \textbf{Patient Facing} & 52\% & (117) & 13\% & (15) & 8.6\% & (41)\\
   Billing for Patient & 2.7\% & (6) & 4.4\% & (5)& 4.6\% & (22)\\
   Website/Portal & 18\% & (41) & 2.7\% & (3)& 1.0\% & (5)\\
   Phone Line & 31\% & (71)& 8.0\% & (9)& 3.1\% & (15)\\
   \midrule
   \textbf{Patient Diversion} & 14\% & (32) & 1.8\% & (2) & 5.9\% & (28)\\
   \midrule
   \textbf{Data Breach} & 0\% & (0) & 6.2\% & (7)& 21\% & (99)\\
   \midrule
   \textbf{Mental Health} & 0\% & (0) & 0\% & (0) & 5.9\% & (28)\\ 
   \bottomrule
\end{tabularx}
\end{center}
\caption{\label{tab:content_count_all_no_change}Number of
    Facebook posts (FB), Facebook comments (FBC), and Reddit posts and comments
    (R) that mention each content category, \textit{excluding} posts and
    comments for the Change Healthcare attack.  A post or comment can appear in
    multiple categories if applicable. The percentages are relative to the total
    number of posts and comments for each platform \textit{excluding} the Change
    Healthcare attack: 226 Facebook posts (FB), 113 Facebook comments (FBC) and
    312 Reddit posts and comments (R).}
\end{table}